%% file: document.tex
\documentclass[sigconf]{acmart}

\author{Christian Amsüss}
\title{Discovery and capabilities of guard proxies for CoRE networks}

\usepackage{colortbl}
\definecolor{cell-red}{RGB}{255, 192, 192}
\definecolor{cell-orange}{RGB}{255, 224, 192}
\definecolor{cell-yelloworange}{RGB}{255, 248, 176}
\definecolor{cell-yellow}{RGB}{255, 255, 160}
\definecolor{cell-green}{RGB}{192, 255, 192}

\usepackage{multirow}

\makeatletter
\renewcommand\paragraph{\@startsection{paragraph}{4}{\parindent}{-.5\baselineskip \@plus -2\p@ \@minus -.2\p@}{-3.5\p@}{\ACM@NRadjust{\@parfont}}}
\makeatother

\begin{document}

\acmYear{2021}\copyrightyear{2021}
\setcopyright{acmlicensed}
\acmConference[DAI-SNAC '21]{Descriptive Approaches to IoT Security, Network, and Application Configuration}{December 7, 2021}{Virtual Event, Germany}
\acmBooktitle{Descriptive Approaches to IoT Security, Network, and Application Configuration (DAI-SNAC '21), December 7, 2021, Virtual Event, Germany}
\acmPrice{15.00}
\acmDOI{10.1145/3488661.3494029}
\acmISBN{978-1-4503-9136-8/21/12}

\include{body}
\end{document}

%% file: body.tex
\begin{abstract}
Constrained RESTful Environments tolerate and even benefit from proxy services. We explore the concept of proxies installed at entry points to constrained networks without any unified management. We sketch proxies of different levels of intrusiveness into applications, their announcement and discovery, and compare their theoretical capabilities in mitigating the effects of undesired traffic that can otherwise exhaust the environment's constrained resources.
\end{abstract}

\maketitle

\section{Introduction}

\input{fig-network.tex}

The Constrained Restful Environments (CoRE) ecosystem
is a family of protocols and usage patterns
for devices that are constrained in several ways, e.\,g. in storage, processing power and energy supply~\cite{rfc7228}.
It is built around the Constrained Application Protocol (CoAP)~\cite{rfc7252},
which enables request/response interactions
directly between constrained devices
over generic Internet connections.

These direct connections are desirable in general due to the wide range of use cases they enable in the Internet of Things (IoT),
but problematic in setups where merely receiving unsolicited messages strains the devices.
For example, a coin cell operated device using the EDHOC key exchange protocol
(Ephemeral Diffie-Hellman over COSE, \cite{draft-ietf-lake-edhoc-12})
can perform the exchange around 50\,000 times before exhausting its battery~\cite{cost-of-oscore} --
even at network speeds of only 1\,kbit/s,
an attack can waste years of battery life in less than a day.

In this short paper we explore to which extent the use of proxies,
i.\,e. network nodes that participate in the CoAP protocol rather than just forwarding IP traffic,
can help protect constrained CoAP devices.
Exploration is done in the form of thought experiments.
Their purpose is to find directions and questions for practical experimentation,
and to spark discussion both on the suitability of the approach
and the requirements on implementations.

To separate desired from undesired traffic,
we will look at ``guard proxies''.
They are CoAP proxies with the purpose to protect a constrained network
or to assist local devices in accessing other thusly guarded devices
(not ruling out additional proxy services provided by them).
They roughly follow the Proxy firewall pattern~\cite{firewall-patterns}.

Guard proxies are described here as being implemented ``on'' routers.
This is only for brevity:
in generalization,
any device can be a guard proxy as long as it is within the local network's security domain
and has unrestricted access to the constrained network.

Even though the devices are constrained,
we assume that they are configurable to the point where they are general purpose devices
within these constraints,
and are set up by their users in ways not necessarily anticipated
by their owner or the local network manager.
Applications such as \cite{draft-ietf-core-dynlink-14} and \cite{micropython} indicate that this is feasible even in constrained devices.

In the course of this text we will look at the protocol components involved,
set up a baseline scenario,
and then explore two forms of guard proxies,
which represent the extremes of the trade-off between intrusiveness into existing systems
and the capability to keep out undesired traffic.
Concluding, we compare the pros and cons of the approaches discussed so far,
and formulate the questions that will shape future work.

\section{Methods}
\label{methods}

Common to all scenarios, we assume that some combination of the following components is used for regular communication.
All these are part of, or being developed for, the IoT ecosystem specified by the IETF (Internet Engineering Task Force) --
either published as a Proposed Standard or active working group items.
The Proposed Standards (CoAP, OSCORE)
are well established for IoT communication
through their inclusion in industry specifications
~\cite{ocf}\cite{lwm2m};
the later items complement the former.


\paragraph{CoAP}
is a web protocol similar to HTTP but tailored for resource constrained devices and networks
~\cite{rfc7252}.
It supports direct communication between devices using UDP
without the need for intermediaries, such as application specific gateways or middle boxes.
It implements patterns of the REST design~\cite{fielding}
like client-server operation, statelessness and the opportunity to do caching in proxies
(voluntarily selected intermediaries),
and can be extended at different layers (e.\,g. by using TCP between less constrained nodes).

\paragraph{OSCORE}
(Object Security for Constrained RESTful Environments)
is a security layer for CoAP that provides confidentiality, integrity, and a strong binding between requests and responses
~\cite{rfc8613}.
It exclusively uses symmetric keys (defaulting to AES-CCM), which necessitates~\cite{rfc4107} the use of a key management protocol.
Its way of building on CoAP facilitates end-to-end encryption even in the presence of proxies,
which is why we choose it here over the more widespread DTLS security layer also available for CoAP.

\paragraph{EDHOC}
is a work-in-progress lightweight key exchange protocol tailored for OSCORE,
and can likewise be transported on top of CoAP
~\cite{draft-ietf-lake-edhoc-12}.
It negotiates an elliptic curve,
performs a Diffie-Hellman key exchange on it,
and provides keys and parameters for OSCORE.

\paragraph{ACE-OSCORE}
is a work-in-progress protocol by which
a trusted third party (the Authorization Server, AS) distributes key material
and indicates the authorization that comes with it
~\cite{draft-ietf-ace-oscore-profile-19};
it is based on the ACE framework (Authentication and Authorization for Constrained Environments, \cite{draft-ietf-ace-oauth-authz-45}).
A client can ask the AS for keys to a server, which the AS provides in a self-contained form for the client to send along to the sever.

\section{Scenarios}
\label{scenarios}

\subsection{Baseline: Indiscriminate network access}
\label{baseline}

We consider a base setup
(depicted in figure \ref{fig-network}a)
of maximally decentralized devices:
communication uses as few centralized services as possible with the selected protocols.
Client and server are both constrained devices (``Class 1'' in the terminology of \cite{rfc7228}),
participating in networks where excessive traffic not only leads to packet loss,
but possibly exhausts the energy budget of the device.
(All later deliberations will apply to setups with a less constrained client as well,
with possible simplifications).

It is assumed that the client discovers the server through a service
to which the server announces its availability, address and any additional metadata:
the rendezvous service (e.\,g. DNS~\cite{rfc6763}, RD~\cite{draft-ietf-core-resource-directory-28}, or plain web links).
When ACE is used, the additional metadata may indicate which AS is used;
if it does not, the AS is assumed to be prearranged.

There is no shared management of the constrained networks.
The list of peers the devices choose to interact with may change over time,
and the peers move throughout the network;
the routers can therefore not know desirable traffic from its address.
(This follows from the assumptions on them being general purpose IoT devices;
in applications where these do not hold, a Manufacturer Usage Description (MUD~\cite{rfc8520}) would address many threats more easily).

\paragraph{The threat model} shared across the scenarios is that
an attacker has learned the address of the server
(e.\,g. by previously having been authorized)
and drains the server's resources by posing as a legitimate client.

The attacker is not assumed to have access to either of the local networks.
The attacker may know addresses of a legitimate client (and spoof its source addresses),
and may be capable of reading (and even replacing) some traffic between the routers.

\paragraph{Threat mitigation}
happens naturally to some extent by the limited bandwidth of the constrained network,
or explicitly by the router throttling incoming traffic.
Seeing CoAP only up to the UDP layer, throttling at a router is limited:
There is no distinction between new and established clients,
and the router can not verify the client's source address.
Thus, throttling at this level affects legitimate and illegitimate accesses alike.
It limits the resources an attacker can drain from the network,
but equally affects legitimate clients.

\subsection{Throttling exemptions for recognized good clients}
\label{exemptions}

\input{fig-revproxy}

For this first scenario, we introduce a guard proxy at the server's network entry point.
(See figure \ref{fig-network}b).

Its proxy operations work the same no matter whether it is a forward or reverse proxy%
\footnote{
As defined in \cite[5.7.2f]{rfc7252}.
Operation of an intercepting proxy is practically possible,
but discouraged in the Internet community~\cite[2.14]{rfc3234}\cite[9.3.1]{rfc3040},
and commonly interferes with the security of non-constrained setups~\cite{https-interception}.
}. This can be implemented in different ways:
The server can (as part of its local network onboarding) know of the services its guard proxy offers,
and announce it as a usable forward proxy at the rendezvous service
using the mechanisms being developed in \cite{draft-amsuess-core-transport-indication-01}.
If the server's application can tolerate being accessed though a reverse proxy
(and the guard proxy provides such),
it can just announce a reverse proxied address instead (thus hiding its network address).
If there is cooperation between the rendezvous service and the guard proxy,
the service may announce a forward or reverse proxy without the server requesting it.
The combination of the latter two is attractive as it enables this scenario without any modification to previous clients and servers,
but requiring such cooperation exceeds our assumptions of independent systems.

Using the announced server information,
clients send requests to the proxy rather than just through the router.
An exemplary exchange is illustrated in figure \ref{fig-revproxy}.

Without further changes, this already has benefits on network performance:
The proxy can manage retransmissions
and adjust message sizes
using parameters adaqeuate for the constrained network,
even when the actual content is hidden in an OSCORE layer.

Throttling as described in \ref{baseline} is still applied,
but done by the guard proxy in a more nuanced fashion.
Traffic that does not go through the guard proxy remains allowed but is throttled more tightly,
still allowing applications not covered by the guard proxy.
This also keeps the system usable if the guard proxy can not be announced at the rendezvous service (and is announced during the key exchange instead).
Such traffic is assigned the lowest priority.
As most legitimate traffic can now be exempt from these policies,
the bandwidth for throttled traffic can be reduced further.

\paragraph{The threat model}
is extended compared to \ref{baseline},
as now the attacker can attempt to force legitimate client traffic
back into throttling.

\paragraph{Threat mitigation:}
The guard proxy has more information available than the router,
and can distinguish between successful and unsuccessful interactions
as well as generally interact on the CoAP level.
Requests from unknown clients are initially suspicious
and subject to severe throttling.
Clients can be informed of the situation~\cite{rfc8516}
and simultaneously be checked for reachability at the claimed address~\cite{draft-ietf-core-echo-request-tag-14}%
\footnote{Prioritizing requests that have been on hold for long is close to violating that document's mandate to ``NOT [...] correlate requests for other purposes than freshness and reachability''; it can be justified by prioritizing requests whose source is known to have been reachable for long time.}.

Requests that are observably authenticated and accepted by the server
can put the sending client on a tentative allow-list,
ensuring that the request and any follow-ups are not subject to throttling.
Of the considered protocols, only OSCORE provides such observability:
The non-OSCORE messages of EDHOC (in the common client-initiated mode) tell the proxy nothing about the client's authorization.
The second message of an ACE-OSCORE exchange does indicate that the server accepts the client's credentials,
but as the token is not regarded as secret, it might be obtained by an attacker.

OSCORE allows and prefers short identifiers.
An attacker that guesses a context identifier that goes with a source / destination address pair
can not bypass the throttling imposed on unauthenticated traffic
(because such a request would not produce a successful response),
but can force legitimate clients off the allow-list.
To mitigate that, the proxy can keep track of the sequence number used with any observed OSCORE context.
For requests with implausibly high sequence numbers,
it can demand that the client demonstrate its reachability again
(and thus keep the attackers traffic off the legitimate client's reputation record);
for sequence numbers already seen on a different token,
it can outright reject them.
Conversely, the same plausibility check can put a mobile client on a higher priority
when a request arrives with a plausible sequence number for a context, but a new address.

This scenario's mitigations do not work against attackers that are capable of reading traffic between networks.
Such an attacker could produce requests with plausible sequence numbers,
or, worse, respond to client requests in such a way that the security context needs to be rekeyed --
not only forcing the client through a throttled operation,
but occupying both sides with costly operations.

%

\subsection{Full guard proxy with token negotiation}
\label{fullguard}

To protect the constrained devices from rekeyings forced by an attacker that can read their traffic,
allow initial key exchanges to be unthrottled,
and at the same time not burden them with additional operations,
we introduce a guard proxy at the client's side,
and let the guard proxies negotiate a tunnel between each other.
(Illustrated in figure \ref{fig-network}c).

The client can discover its guard proxy at onboarding time, like the server does,
and uses it as a generic forward proxy for requests that leave the network.
(This is generally beneficial,
e.\,g. as it decouples retransmissions,
and because it gives an unthrottled path for response traffic if the client's network is under similar protections as the server's).
In the case of an unconstrained client,
the guard proxy's tasks can be performed by the client
with several simplifications.

As the EDHOC or ACE-OSCORE exchanges between client and server are already legitimized by the tunnel,
undesired (or possibly undesired) traffic can be completely blocked from the constrained network rather than throttled.
This entails loss of general Internet connectivity;
it is debatable whether the term ``IoT'' still applies to such setups.

\paragraph{Threat mitigation}
of undesired traffic happens at the server's guard proxy:
It obtains explicit confirmation of the client's legitimacy before burdening the constrained server or network.
Rekeyings between server and client can no longer be forced even by attackers that can read traffic between the routers,
as messages they inject are verified (and their content rejected) by the respective guard proxies.
Such attackers may still interrupt the guards' connection and force them to renegotiate,
but towards the constrained devices this appears as a proxy / network failure at worst.

\input{fig-aceflow.tex}

\paragraph{Setup in the ACE framework}
requires cooperation of all involved parties,
but no previous trust between parties that do not already have trust relationships established,
and no new cryptographic operations other than exchanging more data along established channels.
Figure \ref{fig-aceflow} illustrates this; numbers in parentheses indicate steps illustrated there.

\input{tab-effects.tex}

When the server discovers its guard proxy,
it informs the guard proxy of any AS it is willing to accept clients through (1).
This includes a public key of the relevant AS and the server's ``audience'' identifier.

Both guard proxies provide a public key to their respective devices as part of their discovery (2).

When the server registers its proxy at the rendezvous service,
it includes its guard proxy's public key in the registration's metadata (3).
Similar information is already available to peers connecting to the guard proxy over EDHOC;
publishing it reduces the need for additional exchanges,
and allows leveraging any security mechanisms the rendezvous service provides (e.\,g. \cite[7.3]{draft-ietf-core-resource-directory-28}).
Note that the server does not establish communication with the AS;
this follows the pattern of self-contained tokens of ACE.

When the client contacts the AS to obtain an access token,
it includes both its own guard proxy's public key and the discovered guard proxy's in a new field exclusive to proxy operation (4).
The AS can then associate the authorization to establish communication with the token's audience with the owner of the client's guard proxy's key.

Before the client sends requests to the server through its own guard proxy,
the client shares with it the server address metadata (indicating the public key of the server guard proxy),
as well as the involved AS's address (5).
This step can be skipped if all the relevant metadata was obtained through the client guard proxy
in such a way that the proxy could process and store the information.

Later steps only impose work on the unconstrained parties:
When the client guard proxy eventually receives requests directed at the server (6),
it has all the required information to establish a secure connection to the server's guard proxy.
It knows the AS by address
and has a key that is accepted by it in a unilaterally authenticated connection.
It can request and obtain a token valid for establishing communication with the given audience's guard proxy (7).
When it establishes the connection to the server's guard proxy,
it can use the token to authenticate (8).
The server's guard proxy can verify that the token was issued by an AS indicated by one of its servers,
and allow the client's guard proxy to send requests to the server identified by its ``audience'' identifier.


\paragraph{Setup in EDHOC}
would need more complex interactions of the constrained devices,
as there is no trusted third party to which introduction of the guard proxies can be delegated.
The setup steps would involve either the client issuing a certificate for its guard proxy,
or establishing a secure connection to the server guard proxy on its own.
The server would either need to communicate its acceptance criteria to its guard proxy in full,
or suffer at least some throttled traffic during connection setup
(where intermediate solutions are possible).

These are considered too onerous on the constrained devices to explore further at this point.

\section{Conclusions and Outlook}

The effects expected of the explored scenarios are summarized in table \ref{tab-effects}.
With increasing power to ward off attacks,
the complexity of implementations rises.
The required additions to the constrained side for the full guard scenario
need to compete for ROM, RAM and message exchanges with other applications,
whereas a reverse proxy setup
might be doable for a server in a known network with only a few bytes of added ROM size,
and without changes to the client.

In light of that,
the full guard scenario will need to justify that complexity
by affirming either of three questions:
Does traffic throttled to an acceptable attacker rate slow down key exchanges beyond what applications can tolerate?
Is the ability to set up new connections quickly even under a distributed attack critical to applications?
Is it expected that attackers can read some traffic between systems,
and if so, can the full guard approach provide tangible benefits then?

As these hinge on application specifics,
obtaining input from practically deployed use cases will be necessary.
With that, the behavior of the throttling exemptions scenario can be simulated to assess its suitability.
In parallel, the sketched mechanisms for advertisement and discovery for all scenarios
will need to be specified in sufficient detail that the implementation complexities can be compared.

\bigskip

For either scenario,
a question will need to be answered together with the larger CoRE community:
Are we building the very middle boxes we set out to obsolete with CoAP?

\phantom{
\cite{openclipart}
}

\bibliographystyle{ACM-Reference-Format}
\bibliography{bibliography}

%% file: fig-network.tex
\begin{figure}
%
%
 \includegraphics[width=\columnwidth]{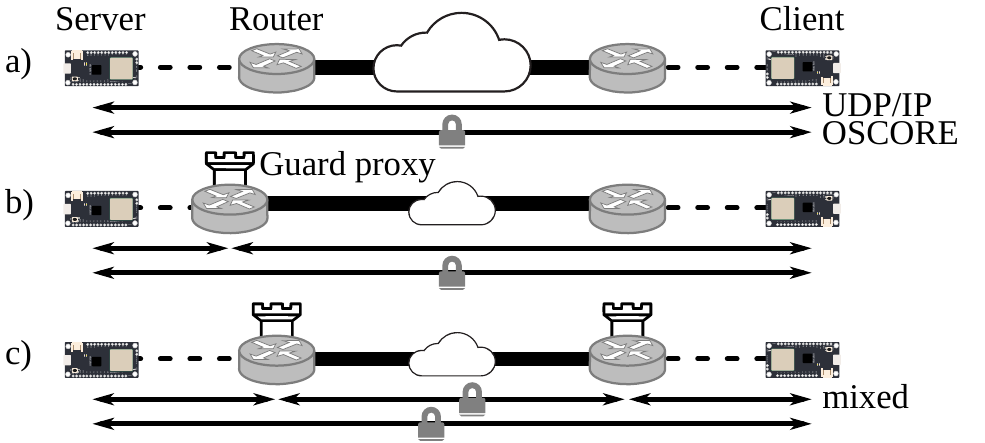}
 \caption{Communication diagrams of scenarios: a) unprotected baseline, b) throttling exemptions, c) full guard proxy. Symbols from \cite{openclipart}.}
\label{fig-network}
\end{figure}

%% file: fig-revproxy.tex
\begin{figure}
%
\includegraphics[width=\columnwidth]{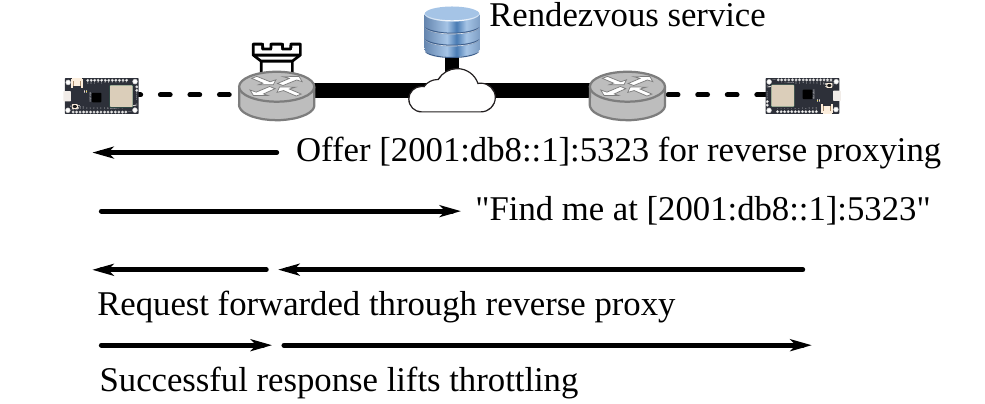}
\caption{Message flow when the server uses the reverse proxy functionality provided by its guard proxy. The client and rendezvous service need no modifications to the baseline setup. Symbols from \cite{openclipart}.}
\label{fig-revproxy}
\end{figure}

%% file: fig-aceflow.tex
\begin{figure}[b]
%
%
%
%
%
%
%
%
\newcommand{\Var}[1]{$\langle#1\rangle$}
  \fbox{
    \begin{minipage}{0.8\columnwidth}
      \raggedright{
        Services of S are best accessed through proxy at address 2001:db8::1, \textbf{which uses \Var{key_{SGP}}}.
      }
     \end{minipage}
  }

 \includegraphics[width=\columnwidth]{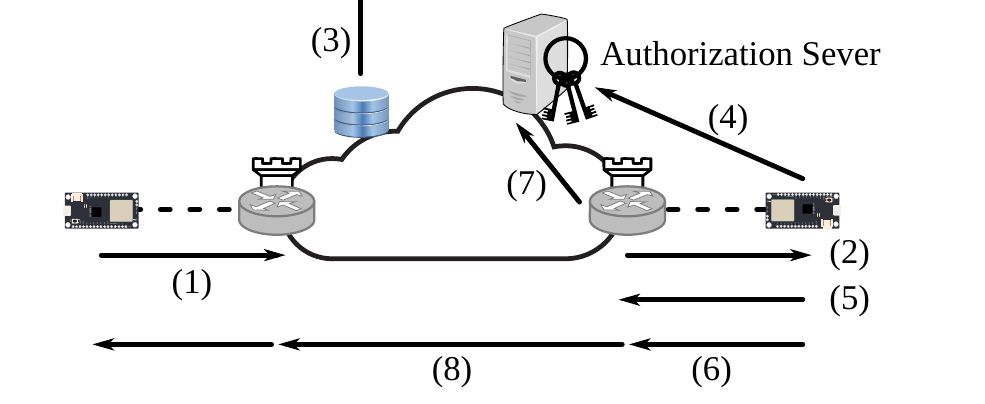}

\raggedright{
(1) ``Proxy for me. \textbf{Accept tokens from \Var{key_{AS}} for \Var{audience}}.''\\
(2) ``I can proxy. \textbf{My key for an AS is \Var{key_{CGP}}}.''\\
(3) Information published at the rendezvous point.\\
(4) ``Token for S please. \textbf{Its guard is \Var{key_{SGP}}, mine is \Var{key_{CGP}}}.''\\
(5) ``You will need this soon: Copy of (3); \textbf{and get keys from \Var{AS}}.''\\
(6) Initial request to S (OSCORE, EDHOC or ACE-OSCORE).\\
(7) \textbf{``I am \Var{key_{CGP}}; a token for \Var{key_{SGP}} please.''}\\
(8) \textbf{Establish context using that token}, then forward (6).
}

\caption{Messages sent to set up ACE for the full guard scenario. Bold text indicates new setup steps compared to setup with client side proxy but no added security. Symbols from \cite{openclipart}.}
\label{fig-aceflow}
\end{figure}

%% file: tab-effects.tex
\begin{table*}

\begin{tabular}{l@{\qquad}|l|c|c|c|c|c|c|c|c|c}
\multicolumn{2}{l|}{} &
    \multicolumn{4}{c|}{ Baseline } &
    \multicolumn{2}{c|}{ \multirow{2}{*}{Exemptions} } &
    \multirow{2}{*}{Full guard} \\
\multicolumn{2}{l|}{} &
    \multicolumn{2}{c|}{ unthrottled } &
    \multicolumn{2}{c|}{ throttled } &
    \multicolumn{2}{c|}{ } &
    \\ \hline
\multicolumn{2}{l|}{Device resource spent under attack} &
    \multicolumn{2}{c|}{\cellcolor{cell-red} high } &
    \multicolumn{2}{c|}{\cellcolor{cell-yellow} low } &
    \multicolumn{2}{c|}{\cellcolor{cell-yellow} low } &
    \cellcolor{cell-green} low or none \\
Behavior of connection setup & under attack &
    \cellcolor{cell-green} good & \cellcolor{cell-red} losses &
    \cellcolor{cell-yellow} throttled & \cellcolor{cell-red} losses &
    \cellcolor{cell-yellow} throttled & \cellcolor{cell-orange} losses$^\dagger$ &
    \cellcolor{cell-green} good \\
Behavior after setup & under attack &
    \cellcolor{cell-green} good & \cellcolor{cell-red} losses &
    \cellcolor{cell-yellow} throttled & \cellcolor{cell-red} losses &
    \cellcolor{cell-green} good & \cellcolor{cell-green} good &
    \cellcolor{cell-green} good \\
\multicolumn{2}{l|}{Implementation effort on constrained devices} &
    \multicolumn{2}{c|}{\cellcolor{cell-green} --- } &
    \multicolumn{2}{c|}{\cellcolor{cell-green} none } &
    \multicolumn{2}{c|}{\cellcolor{cell-yellow} small or none } &
    \cellcolor{cell-orange} medium \\
\multicolumn{2}{l|}{Total implementation effort} &
    \multicolumn{2}{c|}{\cellcolor{cell-green} --- } &
    \multicolumn{2}{c|}{\cellcolor{cell-yellow} small } &
    \multicolumn{2}{c|}{\cellcolor{cell-orange} medium } &
    \cellcolor{cell-red} high \\
\end{tabular}

\caption{Summary of the scenarios: Unthrottled and throttled baseline (\ref{baseline}), with throttling exemptions (\ref{exemptions}; $\dagger$:~Attacker needs to be distributed, otherwise behavior is throttled), and with full guard proxies (\ref{fullguard}). Behavior is ``throttled'' if CoAP interactions complete within an adaequately configured exponential back-off, and with ``losses'' when retransmissions time out.}
\label{tab-effects}
\end{table*}